\documentclass[twocolumn,prl,showpacs,preprintnumbers,amsmath,amssymb]{revtex4-1}
\usepackage{graphicx}
\usepackage{dcolumn}
\usepackage{bm}
\usepackage{longtable}
\usepackage{epsfig}

\begin{document}

\title{$\alpha$-HgS Nanocrystals: Synthesis, Structure and Optical Properties}

\author{A. K. Mahapatra}
\email[Corresponding author.\\]{Email address: amulya@iopb.res.in (A. K. Mahapatra)}
\author{A. K. Dash}
\affiliation{Institute of Physics, Sachivalaya Marg, Bhubaneswar 751005, India}

\begin{abstract}
Well-separated mercury sulfide (HgS) nanocrystals are 
synthesized by a wet chemical route. Transmission electron 
microscopy studies show that nanocrystals are nearly spherical in
shape with average size of 9 nm. Grazing angle X-ray diffraction
confirms that HgS nanocrystals are in cinnabar phase. Particle induced 
X-ray emission and Rutherford back scattering spectrometry analysis 
reveal HgS nanocrystals are stoichiometric and free from foreign impurities.
The optical absorption measurements show two excitonic peaks 
corresponding to electron-heavy hole and electron-light hole transitions, which
are blue shifted by 0.1 and 0.2 eV, respectively, from its bulk value, due to 
quantum size effect. The experimental data obtained by optical absorption 
measurement is simulated with a theoretical model considering the particle 
size distribution as Gaussian.  
 \end{abstract}

\maketitle
\section{Introduction}

     Synthesis of nanocrystals has been of  considerable interest for its
 wide possible application in bio-sensors \cite{alivisatos}, 
catalysis \cite{xu}, light emitting devices \cite{colvin}, 
quantum devices \cite{dabbousi,klein} etc. HgS is also known to be a 
technologically important material for its pronounced dichorism \cite{zallen}, 
birefringence \cite{bond}, photoelectric \cite{kreingol,roberts} and 
acousto-optic properties \cite{sapriel}.
 It is interesting to note that only HgS exists in cinnabar structure 
($\alpha$-HgS) at normal condition among all III-V and II-VI 
compounds \cite{zallen1}. $\alpha$-HgS is a wide band gap 
semiconductor ($E_g$= 2.0 eV). Above $344^0$ C temperature, HgS exhibits a 
zinc-blende modification($\beta$-HgS) and is a narrow band gap 
semi-metal ($E_g$= 0.5 eV).

       Very scarce reports are available on nanocrystalline $\alpha$-HgS although
considerable work has been carried out for other II-VI group semiconductor nanocrystals.
 There are few  reports available on nanocrystalline $\beta$-HgS and also on complex 
 structures like quantum dot quantum well in which a shell of 
 $\beta$-HgS is embedded in CdS quantum dot \cite{qadri,mews,yeh}. Attempts
have been made to synthesize $\alpha$-HgS by Wang {\it et al.} \cite{wang} using 
sonochemical method.  The nanocrystals synthesized in this route are 
of irregular shape, aggregated and hava a wide size distribution.  
		
		 The present work reports a cost effective
wet chemical route to synthesize $\alpha$-HgS nanocrystals in large scale.
The synthesis route is simple as it is a single step process
and carried out in room temperature. Use of polymer during synthesis process 
helps in getting well separated nanocrystals with relatively narrow size distribution. 
The polymer used here is poly vinyl alcohol (PVA). It is a water soluble, transparent 
and high viscous polymer. 

	The size, phase, stoichiometry and purity of the $\alpha$-HgS nanocrystals are 
evaluated by transmission electron microscopy(TEM), Grazing angle X-ray 
diffraction(GXRD), Rutherford back scattering spectrometry(RBS) and 
Particle induced X-ray emission spectrometry(PIXE).  Optical absorption spectroscopy is 
used to estimate the band gap as well as the crystallite size distribution.

\section{Experimental}
 
        The precursors used for the $\alpha$-HgS nanocrystal synthesis are thiourea, 
mercury chloride (HgCl$_2$), ammonia and PVA. 0.001 M aqueous solution of HgCl$_2$ and 
thiourea were prepared separately. Four  aqueous solution of PVA is also prepared. Twenty milliliter of PVA solution is added into 20 ml of HgCl$_2$ solution and stirred for 10 min. Twenty milliliter of thiourea solution is added in it in the same stirring condition. 
Addition of 50 $\mu$l of 3M ammonia solution turns this transparent solution 
into a pale yellow colour solution due to formation $\alpha$-HgS nanocrystals. 
The experiment is done at room temperature (300K).

      The mechanism of formation of HgS nanocrystals in this 
chemical route is supposed to undergo the following steps:\\
\begin{math}
HgCl_2+2NH_4OH  \rightarrow  [Hg(NH_3)_2]Cl_2+2H_2O  \\  
C(NH_2)_2S+NH_4OH  \rightarrow  CH_2N_2+H_2O+HS^{-}+ NH_4 ^{+}  \\
$~$[Hg(NH_3)_2] Cl_2  \rightarrow  [Hg(NH_3)_2]^{2+}+2Cl^{-}  \\
HS^{-}+NH_4OH  \rightarrow  S^{2^-}+H_2O+{NH_4}^{+}  \\
2NH_4^++2Cl^-  \rightarrow  2NH_4Cl  \\
$~$ [Hg(NH_3)_2]^{2^+} + S^{2^-} \rightarrow  HgS+2NH_3\uparrow \\
\end{math}

		TEM is performed using JEOL-2010 operated at 200 KeV electron 
beam energy. For TEM analysis, carbon coated grids are dipped into the 
colloidal solution and held it aloft to dry in ambient. The GXRD is performed 
using Philips  X'Pert system using Cu $K_\alpha$ line as the incident 
radiation. PIXE and RBS measurements are carried out using the 3 MV pelletron 
accelerator with $H^+$ and $He^{++}$ ion beam of energies 2.5 and 3.05 MeV, 
respectively. Optical absorption measurements is carried out by using a 
dual beam Szimadzu  UV-3101 PC spectrophotometer with proper baseline 
correction. For GXRD and Optical absorption measurement colloidal solutions 
are dried on glass substrate. However, for PIXE and RBS colloidal solutions 
are dried on silicon substrate instead of glass substrate because elements present 
in the glass  will be reflected in the spectrum and makes the analysis 
complicated without giving any additional information.

\section{Results and discussion}
\subsection{GXRD and TEM studies}
\begin{figure}
\begin{center}
\includegraphics*[width=9.0cm]{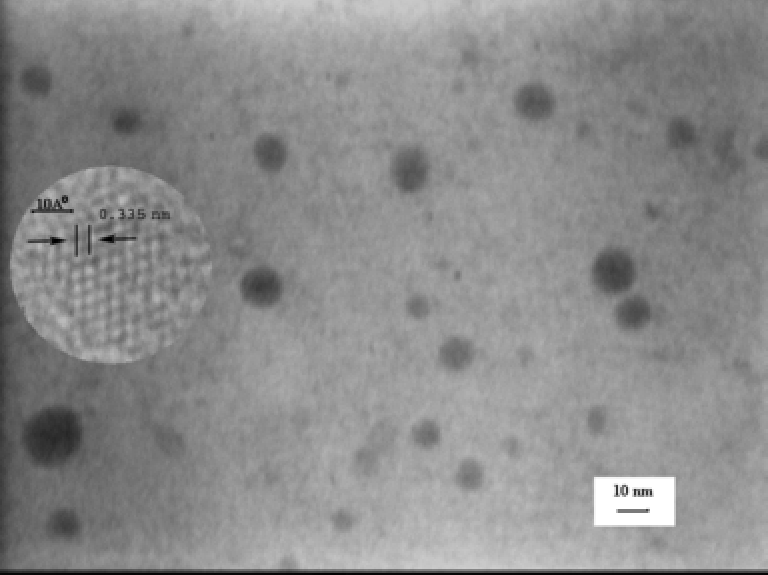}
\caption{TEM micro-graph of the Nanocrystals and HRTEM image is 
shown in the inset.}
\label{tem}
\end{center}
\end{figure}

The formation of HgS nanocrystals is confirmed from TEM. The TEM micrograph of a 
typical HgS nanocrystal sample is shown in the Fig. \ref{tem} and  
high resolution transmission electron microscopy (HRTEM) micrograph
is shown in inset. HRTEM reveals well crystallinity of the nanoparticle. 
The interplanar distance measured to be 0.335 nm as shown in the figure.
This value is same with the interplannar distance between the (1 0 1) planes
of the $\alpha$-HgS and in good agreement with the value estimated from GXRD
and TEM selected area diffraction (SAD) pattern .

\begin{figure}
\begin{center}
\includegraphics*[width=8.0cm]{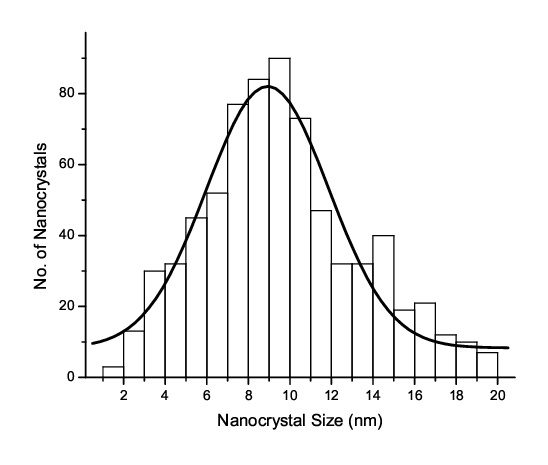}
\caption{Nanocrystal size histogram}
\label{hg}
\end{center}
\end{figure}

  Optical properties of nanocrystals are sensitive to its shape and size. 
 Hence, synthesis of nanocrystals with uniform shape and size, and most 
 importantly accuracy in its measurement, is  essential to derive 
any meaningful conclusion of its optical properties. It can be noted that the 
nanocrystals are nearly spherical in shape and well separated, which helps in finding
the size and its distribution with a better precision. The
size of several nanocrystals ($\approx$700) are recorded. Nanocrystal 
size histogram along with the fitted Gaussian is shown Fig. \ref{hg} .
The particle size  distribution is best fitted with a Gaussian of 9.0 nm
mean and 2.4 nm standard deviation. 

\begin{figure}
\begin{center}
\hspace{-1cm}
\includegraphics*[width=9.2cm]{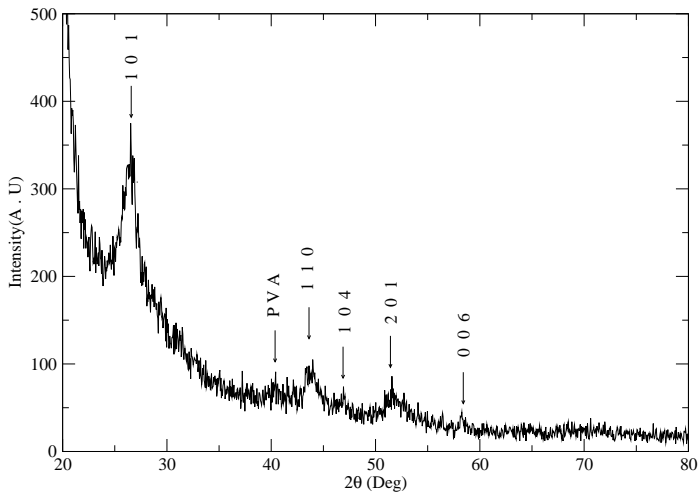}
\caption{GXRD spectrum of $\alpha$-HgS Nanocrystals.}
\label{xrd}
\end{center}
\end{figure}

The average particle size is also calculated by analyzing the (1 0 1) 
peak of GXRD spectra Fig. \ref{xrd}. 
The crystallite size is calculated from Scherrer equation $\it i. e.$ 
 $d$=$\lambda$/$\beta cos(\theta)$. Where $\beta$
is the integral breadth of the diffraction peak, $\lambda$ is the 
wavelength of the incident X-ray and $d$ is the volume weighted 
average crystallite size. The average crystallite size calculated 
using above formula is 7.8 nm which is in the error limit of the value 
obtained from TEM analysis.

\begin{figure}
\begin{center}
\includegraphics*[width=8.0cm]{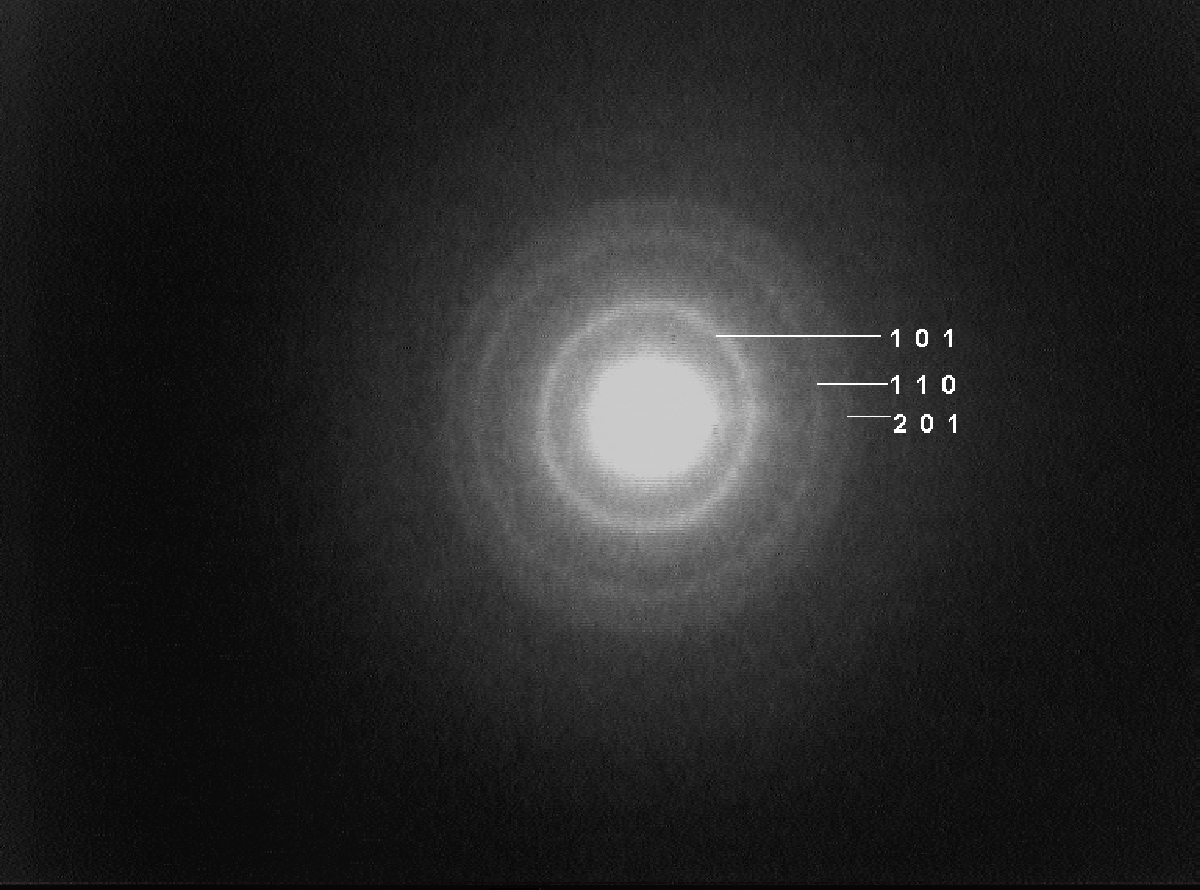}
\caption{Selected area diffraction pattern of $\alpha$-HgS Nanocrystals.}
\label{dp}
\end{center}
\end{figure}

        The interplannar spacing(d) obtained from the GXRD peak 
are in good agreement corresponding to (1 0 1), (1 1 0), (1 0 4), (2 0 1),
(0 0 6) planes of the standard literature data of $\alpha$-HgS 
\cite{JCPDS}. The peak at $2\theta=40.4^0$ in the GXRD 
spectrum is due to PVA\cite{ma}, which plays a major role in avoiding 
aggregation of nanocrystals. The interpalnar spacing($d$) is also calculated 
from the TEM Selected area diffraction pattern and is in good agreement to 
(1 0 1), (1 1 0), (2 0 1) planes. The SAD pattern is shown in Fig. \ref{dp}. 
The continuous ring pattern indicates the polycrystallinity nature of the
sample.  The formula used to calculate $d$-value in the SAD pattern is 
$d_{hkl}$ =$\lambda L$/${R}$. where L is the Camera length, 
R is the Radius of the Ring and $\lambda$ is the wavelength of the electron 
beam. Hence it is concluded that nanocrystals are in cinnabar 
phase ($\alpha$-HgS).  $\alpha$-HgS belongs to 
space group $P{3_1}21({D_3}^4)$ and consists of -S-Hg-S-Hg-S- helical chains, 
six atoms to a turn. The chains are arranged in close packing, so that space lattice is 
hexagonal \cite{zallen1}. The lattice parameters in $a$- and $c$-axis  are 4.149 and 9.495  $\AA$, respectively. It is reported that some materials show a striking 
change in lattice parameters in nano phase \cite{yu, lamber}. However, no such change in 
lattice parameter is observed in the case of nanocrystalline $\alpha$-HgS 
prepared by this chemical route.  

\subsection{PIXE and RBS studies}    
               Presence of any impurity in nanocrystals can 
change its optical properties appreciably. Hence, PIXE and RBS are 
used to identify the composition, stoichiometry and impurity in the 
HgS nanocrystals. 

\begin{figure}
\begin{center}
\hspace{-1cm}
\includegraphics*[width=9.2cm]{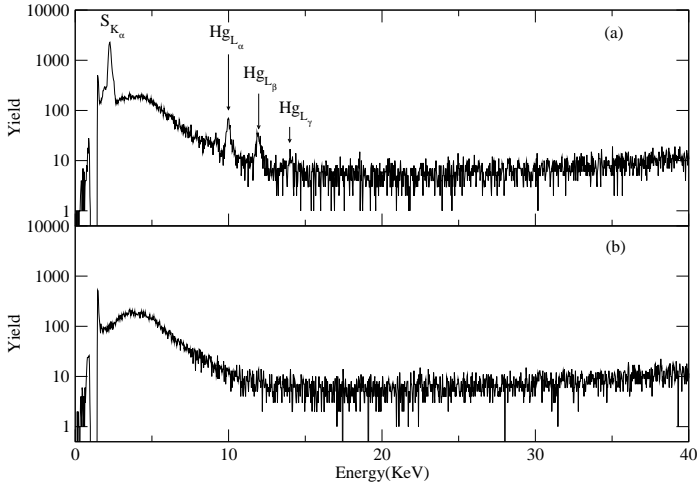}
\caption{PIXE spectra of (a) the HgS nanocrystals on Silicon 
substrate and (b) the Silicon substrate only.}
\label{pixe}
\end{center}
\end{figure}

               In the PIXE experiment, the multiple target holder is
 placed in the plane normal to the beam direction. $H^+$ ion beam of 
energies 2.5 MeV is irradiated to the sample. It knocks out electrons from the 
innermost shells of the atoms and X-rays are emmited with specific energies when 
outer shell electrons change state to inner shell. Si(Li) detector placed at $90^0$ with 
respect to the beam direction in order to detect the x-ray emitted by the de-excitation of the atoms in the sample. The beam current is kept in range of 3-5 nA in order to avoid 
high counting rates at the detector, as it reduces the detection sensitivity due to 
increase of the background noise.  The PIXE spectrum was first calibrated with standard 
sample and then spectra for $\alpha$-HgS nanocrystals on silicon substrate and a bare 
silicon is recorded. For comparison purpose the spectra of  $\alpha$-HgS nanocrystals on 
silicon substrate and a bare silicon is shown in Fig. \ref{pixe}. The intensity ratios and 
energy position of these well resolved peaks are analyzed using GUPIX-2000 software 
\cite{campbell}. Under simulation, it is found that the peaks are coming due to the X-ray 
emmited from different atomic levels of mercury and sulfur. Sulfur is a light element. 
Hence its K X-rays are prominent but mercury being a heavy element, its L X-rays are 
detected. The peaks are marked as HgL$_\alpha$, HgL$_\beta$, HgL$_\gamma$, and 
SK$_\alpha$ as shown in the Fig. \ref{pixe}. Absence of any other peak in the spectra suggests 
that there is no foreign element in the sample even at the ppm level. Here it should be 
noted that detection of element with atomic number less than 11 is limited due to the 
use of Si(Li) detector in the experiment. Hence, presence of element having
atomic number less than 11 in the sample can not be ruled out.  
                 
              In the RBS experiment the multiple target holder is 
placed in the plane normal to the beam direction. The $He^{++}$ ion of energy
3.05 MeV is irradiated to the sample. Some of these ions backscatter due to ellastic 
collision with the atomic nuclei of the sample. The energy of these 
backscattered ions is related to the mass of the target element from which the ion 
backscatters. For the case of heavy target atoms the back scattered energy is high, for 
the case of light target atoms the backscattered energy is low. The energy of the 
backscattered ion is measured by a surface barrier detector, about 20 KeV resolution, 
kept at $150^0$ angle to the incident ion beam. The signals from the detector 
electronic system are in the form of voltage pulses. The height of the pulses are 
propertional to the energy of the backscattered ions falling on it. The pulse height 
analyser stores pulses of a given height in a given voltage bin or channel. The 
spectrum is first recorded for three known standard samples and then $\alpha$-HgS 
nanocrystals dried on silicon substrate is used for RBS data collection under the same 
experimental conditions.  
            
\begin{figure}
\begin{center}
\hspace{-1cm}
\includegraphics*[width=9.2cm]{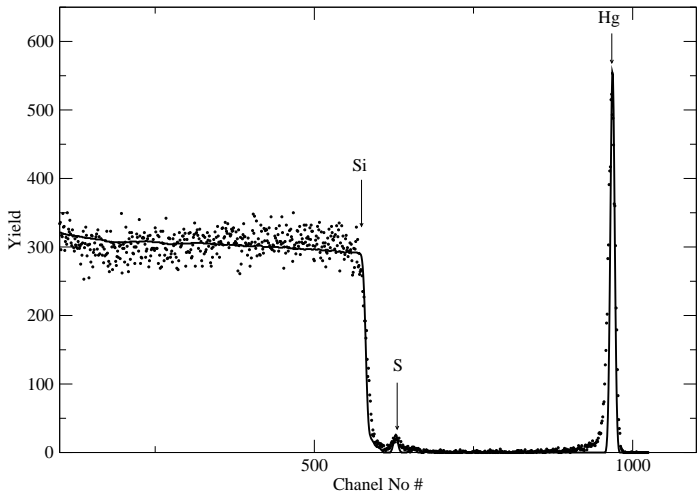}
\caption{RBS spectrum of the HgS nanocrystals on silicon substrate.The solid 
line is the theoretical fit to the experimental data.}
\label{rbs}
\end{center}
\end{figure}

          The experimental data and the simulated profile (using GISA-3 software) is 
shown in Fig. \ref{rbs}. The channel numbers are callibrated in terms of the pulse 
height from the spectrum recorded for three known standard samples. Hence, there is a 
direct relationship between channel number and energy. The prominent peaks in 
the spectrum  are due to backscattering of the $\alpha$-particle from mercury and 
sulfur atom of the HgS nanocrystals. The peaks obtained for mercury and sulfur in the 
spectra have a narrow full width at half maximum suggesting small thickness of the film. The continuous lower energy spectrum is due to the thick silicon substrate. Absence of 
any other peak in the RBS spectra implies no foreign impurity in the sample. 
The thickness of the $\alpha$-HgS thinfilm used for simulation is 20 nm. Under 
simulation, it is found that HgS nanocrystals have stoichiometry.

\subsection{Optical absorption studies}     
\begin{figure}
\begin{center}
\hspace{-0.8cm}
\includegraphics*[width=9.2cm]{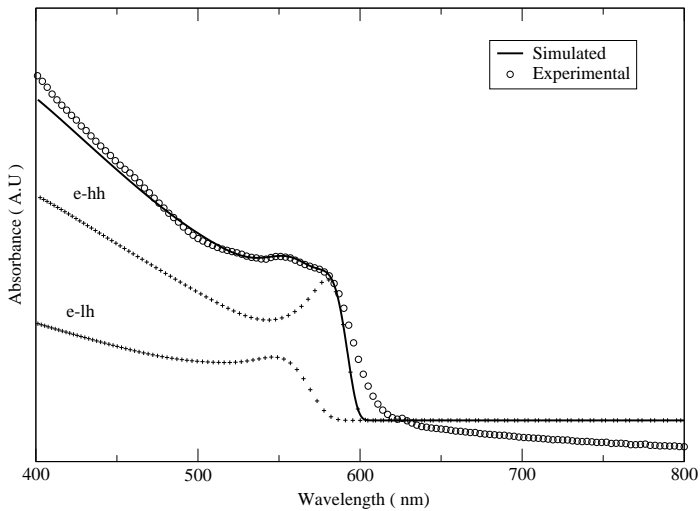}
\caption{Optical absorption spectra of $\alpha$-HgS nanocrystals.Two lower 
curves are contribution due to electron-heavy hole transition and 
electron-Light hole transition.}
\label{optical}
\end{center}
\end{figure}

        The optical absorption spectrum of a zero 
dimensional system is expected to have a series of isolated 
$\delta$-function-like discrete lines. However optical absorption 
spectrophotometer probes a large number of nanocrystals at a time and the
spectrum is the resultant of interaction of all the probed nanocrystals
with light. Hence, the degree of size non-uniformity plays an important role in 
determining the resolvability of the individual peaks corresponding to the 
allowed energy level transitions and the final line shape of the 
spectrum. Guided by this, it is decided to simulate the experimental 
data to find the particle size distribution. The experimental data along with
the simulated profile is shown in Fig. \ref{optical}. 

	Following the work of Wu {\it et al.} \cite{wu} absorption 
coefficient of a large number of cubical nanocrystals of different size 
with a Gaussian distribution can be written as
$$\alpha=\frac{\beta}{a_o}\sum_{n^2}\frac{g(n^2)}{\xi n^2}e^{-
{({n}/{x}-1)^2}/{2\xi^2}}$$

	where two dimensionless parameters,reduced photon energy($x^2$) and 
relative standard deviation of the nanocrystal size ($\xi^2$) is defined as
follows:

	$$x^2=\frac{\hbar\omega-E_g}{\frac{\pi^2\hbar^2}{2\mu{a_o}^2}}$$ and
      $$\xi=\frac{D}{a_o}$$

Here $E_g$ is the bulk band gap,$\mu$ is the reduced mass of the 
electron-hole pair, D is the Standard deviation of the Gaussian distribution 
and $\beta$ is the optical transition dependent constant. Nanocrystals assumed
cubical and  $a_o$ is the average side length of the nanocrystals. However,
$\alpha$-HgS nanocrystals are nearly spherical as observed from TEM measurement. Hence 
reduced photon energy is  modified as 

	$$x^2=\frac{\hbar\omega-E_g}{\frac{2\hbar^2}{\mu{d}^2}}$$ and
 	$$\xi=\frac{D}{d},$$

where $d$ is average diameter of the spherical nanocrystals.  

	The energy levels $\it{n}$ are determined by the roots of 
the spherical Bessel functions $\chi_{ml}$ with $\it{m}$ being the 
number of the roots and $\it{l}$ being the order of the function. $g(n^2)$ 
is the degeneracy in energy level $\chi_{ml} $ and takes care the intensity
contribution due to it.
 
	     The resolvability of peaks and the final 
line shape of the absorption spectrum of a quantum dot system is 
affected due to the contribution from both electron-heavy 
hole(e-hh) and electron-light hole(e-lh) transitions. The average particle 
size is 9.0 nm, as obtained from the TEM analysis,and Bulk band gap is
2.0 eV \cite{zallen1}. A survey of literature shows a lack of satisfactory 
data  on the effective mass of the electron and hole of $\alpha$-HgS. Hence the
parameters used for simulation are $\mu_{e-hh}$, $\mu_{e-lh}$, relative 
intensity of contribution due to electron-heavy hole and electron-light
hole (I) and D. The experimental data best fits for $\mu_{e-hh}$ = 0.14$m_0$,
$\mu_{e-lh}$ = 0.07$m_0$, I = 1.4 and D = 1.7 nm. Hence nanocrystal size estimated 
to be 9.0 $\pm$ 1.7 nm. The standard deviation obtained under simulation is 
nearer but less than the value obtained from the TEM analysis. It should be 
noted that the number of nanocrystals probed under optical absorption 
spectroscopy is much larger than the number of nanoparticles analyzed under TEM. 
The excitonic peak corresponding to electron-heavy hole and electron-light 
hole transition are blue shifted 0.1 and 0.2 eV, respectively, from its bulk 
value due to quantum size effect.

\section{Conclusion}
         A low cost synthesis process of  well-separated  $\alpha$-HgS
nanocrystals using the chemical route is presented. Average size of the 
nanocrystals are 9.0 nm as determined from the TEM analysis. $\alpha$-HgS 
Nanocrystals are pure and stoichiometric. The theoretical model fits well to the
experimental data of optical absorption spectroscopy and suggests a narrow size 
distribution of nanocrystals.

\section*{Acknowledgment}
The help and constant encouragement received from Dr. S. N. Sahu, 
Dr. R. K. Choudhury and Dr. P. V. Satyam, Institute of Physics, are gratefully 
acknowledged.

\end{document}